\documentstyle[preprint,aps]{revtex}

\begin{document}

\draft


\title{
\begin{flushright}
{\rm Oslo-TP-5-94,
gr-qc/9404058}
\end{flushright}
Aspects of
the Thermo-Dynamics of a Black Hole\\
with a Topological Defect}


\author{Bj\o rn Jensen \footnote{Electronic address: BJensen@boson.uio.no}}

\address{
Institute of Physics, University of Oslo,
P.O. Box 1048, N-0316 Blindern, Oslo 3, Norway}

\date{April 28, 1994}

\bibliographystyle{unsrt}

\maketitle


\begin{abstract}
Aspects of the thermo-dynamics of a black hole which
is either pierced by a
cosmic gauge string or contains a global monopole are investigated.
We also make some comments on the physical significance
of the fact that the gravitational mass carried by
a global monopole is negative. We note in particular
that the negative monopole mass implies a gravitational
super-radiance effect.\\

\smallskip

\end{abstract}


%
%

\section{Introduction}

According to most fundamental particle theories,
phase transitions have occurred in
the early universe. Such transitions
may have created various kinds of topological defects
such as domain walls, cosmic strings, monopoles and
textures \cite{Vilenkin1}.
These objects are believed to
have contributed to the creation of
some of the large scale structures found in the
observable part of the
universe today. It has further been pointed out
that topological defects can be sources
for (topological) inflation \cite{Linde2,Linde3}. Hence these
objects may also shape the universe on very
large scale and give rise to an effective fractal geometry
\cite{Linde2}.\\

\smallskip

In this article we will be concerned with some
effects produced locally by a cosmic gauge string
and a global monopole. We will more specifically
be interested in situations where a cosmic string pierce
through a black hole and when a global
monopole has been swallowed by a black hole \cite{Aryal,Lousto,Jing,Yu}.
It has recently been shown that the rate of change of the
energy $E$ of
a black hole which exhibits a topological defect in
this way and the rate of change of the
natural mass $M$ assigned
to such a structure do not coincide \cite{Aryal,Jing,Yu}.
On this ground it is argued that the first law
of black hole thermo-dynamics $dM=TdS$, where
 $T$ is the temperature of the
black hole horizon and $S$ its
entropy has to be modified
when a black hole exhibits a topological defect \cite{Jing,Yu}.
The primary aim of this work is to show that this
conclusion is misleading. We show that the energy flowing into
a black hole with a topological defect ($dE$) naturally
equals the change in the gravitational mass of the whole structure,
i.e. $dE=dM c^2$. From the general thermo-dynamical relation
\begin{equation}
dS=\frac{dE}{T}
\end{equation}
it then follows that the entropy
carries a similar functional relation to the gravitating energy
of the black hole structure as in the Schwarzschild
solution $S=4\pi\alpha^{-2}M^2$ where $\alpha^2$ is a parameter
that characterize the topological defect.\\

\smallskip

We will also consider some
general aspects of the
gravitational properties
of a single global monopole.
Of particular concern will be the fact that
a global monopole naturally can be assigned a
negative gravitating mass.
It is shown that due to its negative mass
a global monopole will
give rise to a
gravitational super-radiance effect.
It follows that global monopoles are
gravitationally unstable objects.

\section{The concept of energy}

In pure Einstein theory it has been proved that the total
energy (the ADM mass) carried by an isolated system, i.e. one that
generates an
asymptotic Minkowski geometry, is positive \cite{Schoen}. Due to the essential
role
played by the asymptotic condition this
theorem can not be carried over to solutions of Einsteins
theory which does not display a Minkowskian
asymptotic structure. This
problem also arise in Kaluza-Klein and super-string
theories where a system can have negative energy even
when the asymptotic behavior approach $M_4\times K$ where
$M_4$ is the four-dimensional Minkowski space and
$K$ is a compact internal space \cite{Brill,Corley}.

In a neighborhood of space-like infinity a black hole
pierced by a cosmic gauge string
or a black hole which has swallowed a global monopole gives rise
to a non-Minkowskian structure. This structure arises since
 strings and monopoles effectively cuts out either a deficit angle
(strings)
or a deficit cone (monopoles)
 from the geometry compared to corresponding space-times
without topological defects of this kind.
This means that if one is to measure the
surface area of a sphere $S^2$ with
a physical radius $r$
 with a monopole
in the
center the area will read $A=\alpha^2 A(S^2)$
with $\alpha <1$. Here $A(S^2)$ denotes
the area of a similar sphere in the Minkowski geometry and $\alpha$
is a constant that characterizes the energy content of the defect.
The appearence of an $\alpha$-factor in the angular part
is
the only difference compared to the Minkowski metric
in the asymptotic region.
This means in particular
that redshift of photon energies for photons propagating
from a region near a string or a
monopole or a black hole with a defect
to infinity still vanishes there. This allow us to define
a meaningful mass concept in these space-times
along similar lines as in asymptotically
flat space-times.

Consider a black hole with a topological defect.
Let $\xi^a$ be a time translation Killing vector
field which is
time-like near infinity such that
$\xi^a\xi_a=-1$
and which has vanishing
norm on the event-horizon. Let $\vec{N}$ be a second Killing vector field
orthogonal to the event-horizon with normalization such
that $N^aN_a=1$ near infinity.
Let $S$ denote the region outside the
event-horizon of the black hole. The boundary $\partial S$
of $S$ is taken to be the event-horizon $\partial B$
and a two-surface $\partial S_\infty$
at infinity. It then follows that the mass $M^\infty$ inside $\partial
S_\infty$
as measured by a static observer at infinity can
be deduced from \cite{Bardeen}
\begin{eqnarray}
\int_{\partial B+\partial S^{\infty}}\xi^{a;b}d\Sigma_{ab}=
-\int_{S}R^a\, _b \xi^b d\Sigma_a \, .
\end{eqnarray}
$R^a\, _b$ denotes the Ricci tensor.
The integral over $\partial B$ is the surface
gravity $\kappa =N_b\xi^a\nabla_a\xi^b$ multiplied with the
surface area $A$ of $\partial B$.

\section{A black hole with a topological defect}

The possible presence of global monopoles is
indicated by the non-triviality of $\pi_2(M)$ (i.e. $\neq I$) where
$M$ is the group manifold of the unbroken global
symmetry group of the theory.
To be specific we will consider a theory with a
Lagrangian on the form (we work in units such that $c=\hbar =k_B =G=1$)\
\begin{equation}
L=\frac{1}{2}g^{\mu\nu}\partial_\mu\phi^a\partial_\nu\phi_a-
\frac{1}{4}\lambda \left(\phi^a\phi_a-\eta^2\right)^2\, .
\end{equation}
Here $\phi^a$ ($a=1,2,3$) is a
triplet of scalar fields with global $O(3)$
symmetry which is spontaneously
broken down to $U(1)$. Due to the this
surviving unbroken symmetry this
model will contain solitons (global monopoles
or hedgehogs)
which carry $U(1)$ charges with masses of
the order of the scale of the symmetry
break-down ($\sim 10^{16}$GeV). In Schwarzschild coordinates
the metric field outside the
core region of a global monopole
is to a high degree of
approximation given by \cite{Barriola}
\begin{eqnarray}
ds^2&=&-\left(\alpha^2 -\frac{2m}{\rho}\right)d\tau^2+
\left(\alpha^2 -\frac{2m}{\rho}\right)^{-1}d\rho^2+
\rho^2d\Omega^2 _2.
\end{eqnarray}
$d\Omega^2 _2$ represents the infinitesimal surface element
on the unit sphere and the radius $\delta$ of the core
region
is given by
$\delta\sim 2(\lambda\eta^2 )^{-1/2}$.
The parameters $m$ and $\alpha$ in the line-element are similarly
\cite{Harari}
\begin{equation}
m\approx -\frac{16\pi}{3}\frac{\eta}{\sqrt{\lambda}}\, ,\,
\alpha^2=1-8\pi\eta^2\, .
\label{approximatemass}
\end{equation}
This geometry is
not asymptotically flat. In order to see this we
perform the coordinate transformation
$\tau\rightarrow t=\alpha\tau\, ,\, \rho\rightarrow
 r=\alpha^{-1}\rho$
along with the replacement of $m$ by $M=\alpha^{-3}m$.
The metric then takes the form
\begin{eqnarray}
ds^2=-(1-\frac{2M}{r})dt^2+(1-\frac{2M}{r})^{-1}dr^2+
\alpha^2r^2d\Omega^2 _2\, .
\end{eqnarray}
At infinity
this structure approximates to
\begin{equation}
ds^2=-dt^2+dr^2+\alpha^2r^2 d\Omega^2 _2\,
\end{equation}
which is trivially seen to describe a curved space.
Compared to Minkowski space-time a solid cone is
missing from the space when $\alpha <1$. We will
always assume that this condition is fulfilled.

A geometry very similar to the geometry generated by
a global monopole (or a black hole with a global
monopole) is found when a
cosmic gauge string pierce through
a black hole. The metric of such a structure is given by \cite{Aryal}
\begin{equation}
ds^2=-(1-\frac{2M}{r})dt^2+(1-\frac{2M}{r})^{-1}dr^2+
r^2(d\theta^2+b^2\sin^2\theta d\phi^2)\, .
\end{equation}
$b$ is a constant assumed less than unity.
It follows that this geometry has properties at infinity
that resembles the effects produced by a monopole. We will
in the following therefore mainly concentrate
our attention to the situation when
a black hole has captured a global monopole.\\

\smallskip

The reason for the curved asymptotic structure
in the global monopole space-time
is that even though the triplet of scalar fields
define a state of lowest energy outside the
monopole core it is not a state of zero energy,
i.e. the energy-momentum tensor have non-zero
$T\tau\tau$ and $T\rho\rho$ components \cite{Harari}. This
tensor is furthermore Lorentz-invariant in the
radial direction. A calculation of the gravitational
mass $M^v$ outside the core-region from
eq.(2) reveals that this
feature of the energy-momentum tensor results in
$M^v=0$ since $T^\tau\, _\tau -T^\rho\, _\rho$
vanishes identically. The true vacuum outside the monopole will
therefore not give rise to an acceleration field
felt by an observer initially at rest relative to the global frame.
Of-course the presence of this vacuum structure do
exert a residual gravitational action since two
initially parallel light rays passing on opposite sides
of the monopole will converge.
This vacuum structure does seemingly also have cosmological
implications. Let $R$ denote the expansion factor of the universe
in the Robertson-Walker line-element. The ``passive''
mass $M_p$ of the monopole field outside the core
of the monopole inside a volume very much larger
than the volume occupied by the core is approximately
given by $M_p\sim \alpha^2\eta^2r$ \cite{Harari}.
The equation governing the Robertson-Walker line-element then becomes
\begin{equation}
\frac{1}{2}\dot{R}^2-4\pi\frac{\rho r^3}{r}=-\frac{k}{2}
\end{equation}
where $k$ is the effective curvature of the universe. With
$M_p$ used as a measure of $\rho r^3$, i.e. $\rho r^3\sim M_p$,
we see that the monopole field will give rise to an
effective curvature $k_{eff}$
 of the universe $k_{eff}=k-2\pi\alpha^2\eta^2$.
It is unclear whether this will have any significant physical
effects.\\

\smallskip

The core-region of a topological defect is
assumed to reside in a state corresponding to the false
vacuum state, i.e. the state of lowest energy before the
symmetry break-down in the theory.
It is found in particular that the monopole core can be modeled by
a sphere with
$T_{\mu\nu}=-\rho \mbox{diag}(-1,1,1,1)$ \cite{Harari}.
It follows that the Tolman mass analog from eq.(2)
 in the core-region is negative.
Measured from
infinity the global monopole configuration
will therefore
naturally be assigned a negative inertial mass
which is reflected in the negativity of $m$.
It also follows that we should not
expect such an object to be stable since we would
expect that the monopole configuration will seek towards
ever decreasing energy states whenever possible. This follows
since the energy steaming from a simple constant re-scaling
of the energy ($g_{\mu\nu}\rightarrow \mu^2 g_{\mu\nu}\Rightarrow
E\rightarrow \mu E$) of a negative energy object does not have a
lower bound as is the case when the energy
of the object is positive.

To give another
 argument in favor of this view we
consider a scalar
field $\psi$ propagating in the geometry eq.(6).
As in the Schwarzschild geometry
we expect that the
 dynamical development of metric perturbations resembles
the dynamics of $\psi$. In this case the equation
govering the evolution of $\psi$ can be brought to the form
\begin{equation}
\partial_t^2\psi =\partial_{r_*}^2\psi -V(r_*)\psi
\end{equation}
where $r_*$ is as usual a
Regge-Wheeler coordinate defined such that
 $r=2M$
is moved to $r_*=-\infty$. It then follows that
the operator $A$ defined by
\begin{equation}
A=-\frac{d^2}{dr_*^2}+V(r_*)
\end{equation}
is a positive self-adjoint
operator on the Hilbert space ${\cal L}^2(r_*)$ of square
integrable functions of $r_*$. Multiply the wave equation
with $\partial_t \psi^*$
(where $\psi^*$ is the complex conjugate of $\psi$)
and integrate over $r_*$. We then
obtain
\begin{equation}
\int |\partial_t \psi |^2dr_*+\int \psi^*A\psi dr_*=C
\end{equation}
where $C$ is a constant
which can be taken positive. Since the last integral is
positiv due to the positivity of $A$ in the Schwarzschild
geometry $C$ will represent an upper bound for the
first integral. In particular uniform exponential growth
in time of initially well behaved perturbations is
ruled out. However, in the global monopole geometry
it is easily seen that $A$ is no longer positive since
 for modes with
sufficiently small angular momentum $V(r_*)$ is negative.
In Schwarzschild coordinates eq.(6) the potential reads
\begin{equation}
\alpha r^3V(r)=(r+2|M|)(l(l+1)-2\alpha |M| r^{-1})
\end{equation}
where $l$ is the usual angular momentum quantum number.
Note that the potential is negative definite for the
 zero angular momentum mode. In general the
potential will be negative whenever
\begin{equation}
r l(l+1)< 2\alpha |M|\sim \frac{8\pi\eta^2}{3}\delta\, .
\end{equation}

The non-existence of an upper bound for $\psi$ indicates
that incident radiation may be
scattered with a total cross-section that exceeds
unity, i.e. more energy scatters back to infinity than
exhibited by the incident radiation. This can be seen
as follows \footnote{A detailed investigation will be
presented elsewhere.}. To the field $\psi$
with an energy-momentum tensor $T^\mu\, _\alpha$
we can assign a
conserved canonical
current $j^\mu =\xi^\alpha T^\mu\, _\alpha$ relative to the Killing observer.
Let this observer monitor the global monopole
in the region between an
initial space-like surface $S_0$ to a final space-like surface
$S$ in the future of $S_0$.
 Assume that the monopole is contained in a spherical
shell with radius $r$ and that the space-like hyper-surfaces
between $S_0$ and $S$ can be parameterized by an
affine parameter $\lambda$.
{}From the conservation equation $\nabla_\mu j^\mu =0$ it then follows
that the difference between the energy within the shell in
the surface $S_\lambda$, $E_\lambda$, and the corresponding
energy in $S_0$, $E_0$, is given by
\begin{equation}
E_\lambda
-E_0=-\mbox{lim}\!\!\!\!\!\!\!\!\!\!\!\raisebox{-1.9ex}{$r\rightarrow\infty$}\;\int_{0} ^{\lambda}\xi^\alpha T^\mu\, _\alpha
N_\mu r^2d\Omega d\lambda =-E_{\mbox{rad}}\, .
\end{equation}
$E_{\mbox{rad}}$ denotes the total radiated energy in
the region between $S_0$ and $S$. With the boundary condition that
$E_{\mbox{rad}}\geq 0$ it follows that
when $E_0$ is positive
then $E_\lambda$ either
equalls or is less than
$E_0$. If on the other hand $E_0$ is negative then $E_\lambda <0$ and
$|E_\lambda |$ will be larger than the initial energy. In $S_0$ lett
a time-dependent $l=0$ mode exist such that no positive energy
passes through the surface at $r$ in the negative radial direction.
Since no upper bound exists for $\psi$ and on the assumtion that
no time-dependent bound states exist the system must radiate energy to
infinity. It follows that $E_\lambda <0$ and energy is effectively extracted
from the system. This is a gravitational analog to the well
known super-radiance phenomenon in the scattering
of charged bosons off large electric potentials. The
waves scattered off the monopole
 will carry a net positive energy to infinity
which must be compensated by letting the Tolman mass
analog of the monopole grow
even more negative. The situation is somewhat similar to
the situation when particles are scattered off
an isolated rotating star or a rotating
black hole. When the star rotates sufficiently fast
an ergo-region develops outside the outer surface of the
star. Particles coming in from infinity with certain energies
(the super-radiant modes) can be scattered off the star and
back to infinity with an energy larger than their
in-coming energy \cite{Friedman}. A finite amount of energy
(the rotational energy) can
be extracted from the star in this way. However, in the case of the
global monopole an upper bound for this energy does not
present itself in a similar way. A similar phenomenon does not
occur when particles are scattered off cosmic gauge strings since
these carry a vanishing gravitational mass.\\

\smallskip

We now consider the situation when a black hole has
swallowed a global monopole. The metric of such a
configuration is still assumed
 given by eq.(4) but with $m$ taken
positive in order to conform with the cosmic censorship
hypothesis.  Let $\vec{\xi}=\alpha^{-1}\partial_\tau$ and
$\vec{N}=\alpha\partial_\rho$. It follows that $\kappa =
\alpha^3 (4m)^{-1}=(4M)^{-1}$ and $A=
16\pi \alpha^{-4}m^2=16\pi\alpha^2 M^2$.
The mass $M^\infty$ of the black hole is then given by
\begin{equation}
M^\infty =\alpha^{-1}m =\alpha^2M\, .
\end{equation}
Note that neither $m$ nor $M$ has the status as graviational mass alone.
An observer at infinity will thus naturally assign
an inertial energy $E=M^\infty$ to the black hole.
By the use of eq.(1) and $\kappa =2\pi T$
\footnote{This relation still holds since the
Eucledian section of the Wick rotated analytic extension of eq.(6)
is independent of $\alpha$.} it then follows that
the entropy of the black hole configuration
is given via
$dS=4\pi\alpha^{-2} d(M^\infty )^2$,
i.e. $S=4\pi\alpha^{-2} (M^\infty )^2$ (up to
an additive constant). This relation is rather
important since it relates the black hole entropy directly
to the gravitating energy ``content''
of the black hole. Also note that these results are
independent of the coordinate system employed. It is furthermore clear
that the usual relation between $S$ and event-horizon area holds,
i.e. $S=\frac{1}{4}A$.\\

\smallskip

Relative to the geometry eq.(6) it follows
that the quasi-newtonian potential $\Phi$ is given by
$\Phi =-Mr^{-1}$. The corresponding gravitational
acceleration $g$ is $g=|\nabla\Phi |=Mr^{-2}$. Let
Newton's gravitational law
take the standard form
$\nabla^2\Phi =\gamma\rho$ where $\gamma$ is the coupling
constant and $\rho$ the mass density. Integrate this equation
over the volume $V$ inside the spherical surface $S$ naturally defined by
the observer at infinty
\begin{equation}
\int_S\nabla\Phi\cdot d\vec{S}=\gamma\int_V\rho dV\equiv\gamma M^\infty\, .
\end{equation}
The gravitational acceleration is then given by
$g=\gamma M^\infty (4\pi \alpha^2 r^2)^{-1}$
where $r$ is the radial coordinate at which $S$ is positioned.
Equality between the two expressions for $g$ is achieved by
defining $\gamma\equiv 4\pi$. Hence, again we find that
$M^\infty =\alpha^2 M$.\\

\smallskip

When the black hole either emits radiation or
swallows mass it is primary $M^\infty$ that
is expected to change.
If one assumes that the mass of the hole is
changing slowly one can write $m(\tau )=m_0+\dot{m}\tau$ where
$m_0$ and $\dot{m}$ are constants such that $\dot{m}<<1$. The
energy flow in the radial direction
relative to the global coordinatization
to first order in $\dot{m}$ then becomes
$T_\tau\, ^\rho =2\dot{m}\rho^{-2}$ \cite{Jing}. It follows that
the total energy $\dot{E}$ that flow through a surface $\Sigma$
relative to the $\xi^\mu$ observer
per coordinate time to
first order in $\dot{m}$ becomes
\begin{equation}
\dot{E}=\int\xi^\mu T_\mu\, ^\nu d\Sigma_\nu =
 \int\xi^\tau\alpha^2 T_\tau\, ^\rho d\Sigma_\rho =
\dot{m}\, .
\end{equation}
The energy-flow that is related to
the change in gravitational mass is a change in energy per unit {\em proper}
time $\hat{E}$. If we assume that we perform our measurements of
the energy-flow in the asymptotic region it follows
that the change in the gravitational mass of the black hole
with a monopole asymptotically approaches $\hat{E}=\alpha^{-1}\dot{m}$. This
is in exact correspondence with eq.(16) since
$\hat{E}$ scales as $M^\infty$, i.e. $\hat{E}$
is naturally interpreted as
the change in $M^\infty$ per unit proper time.
This discussion carry almost unaltered over to the situation
when a cosmic gauge string pierce through a black hole
(replace $\alpha^{2}$ with $b$).\\

\smallskip

\section{Discussion}
In this work
we noted that due to the negative energy carried by
a global monopole
the condition for the occurence of
super-radiance is satisfied
when particles are
scattered off such an object. It implies
that the monopole apparently may radiate
an inifinte amount of energy to infinity
\footnote{Of-course at some point
back-reaction effects must be taken into
account.}. We also found that
when the energy measure was carefully
defined no anomalies in the laws of black hole thermo-dynamics
of a black hole which has captured a monopole
results as indicated in \cite{Jing,Yu}.

Systems containing black holes and objects with negative
energy are certainly very interesting. When a black hole
swallows a global monopole the mass of the resulting hole
must decrease. This does not break with Hawkings
area theorem for black holes since a global monopole breaks
the strong energy condition
due to its negative mass. What seems more interesting is that
the entropy as derived in this work is determined not only
by $M^\infty$, as usual, but also by $\alpha$. Even though
$M^\infty$ of the black hole
 decreases when a hole captures a monopole we must
also take into account the effects produced by the appearence
of $\alpha$ in $S$. It might be that $dS\geq 0$ when
one considers such a process since the $\alpha$-factor is assumed
less than unity.

\section{Acknowledgments}

It is a pleasure to thank H.H. Soleng for many
discussions on the properties of global monopoles
and C.O. Lousto for usefull comments and for calling my
attention to ref.(5).
I thank
NORDITA for hospitality
during the time which parts of this
work was carried out
and I acknowledge
the Norwegian Research Council (NFR)
for a travelling grant, Project No.:420.94/017 .


\end{document}